\documentclass[a4paper,11pt]{article}
\usepackage{pos}

\newcommand{\OpenLoops}{{\rmfamily\scshape OpenLoops}}

\newcommand{\Collier}{{\rmfamily\scshape Collier}}
\newcommand{\OneLoop}{{\rmfamily\scshape OneLoop}}
\newcommand{\Mathematica}{{\rmfamily\scshape Mathematica}}
\newcommand{\Fortran}{{\rmfamily\scshape Fortran}}
\newcommand{\FIRE}{{\rmfamily\scshape FIRE}}
\newcommand{\FIESTA}{{\rmfamily\scshape FIESTA}}
\newcommand{\msbar}{\overline{\mathrm{MS}}}

\newcommand{\reffi}[1]{\mbox{Fig.~\ref{#1}}}

\newcommand{\refse}[1]{\mbox{Section~\ref{#1}}}

\newcommand{\f}[2]{\frac{#1}{#2}}

\newcommand{\ssst}[1]{\scriptscriptstyle{\text{#1}}}
\newcommand{\nosss}[1]{#1}

\newcommand{\bea}{\begin{eqnarray}}
\newcommand{\eea}{\end{eqnarray}}
\newcommand{\be}{\begin{equation}}
\newcommand{\ee}{\end{equation}}
\newcommand{\ba}{\begin{align}}
\newcommand{\ea}{\end{align}}
\newcommand{\beas}{\begin{eqnarray*}}
\newcommand{\eeas}{\end{eqnarray*}}
\newcommand{\bes}{\begin{equation*}}
\newcommand{\ees}{\end{equation*}}
\newcommand{\bas}{\begin{align*}}
\newcommand{\eas}{\end{align*}}

\newcommand{\eps}{{\varepsilon}}

\newcommand{\lb}{\left(}
\newcommand{\rb}{\right)}

\newcommand{\idop}{1\!\!1}

\newcommand{\momk}[1]{k_{\nosss{#1}}}

\newcommand{\mass}[1]{m_{\nosss{#1}}}
\newcommand{\heli}{h}

\newcommand{\helicheck}{\check h}

\newcommand{\helihat}{\hat h}

\newcommand{\momq}{\bar{q}}

\newcommand{\calR}{\mathcal{R}}

\newcommand{\calC}{\mathcal{C}}

\newcommand{\calK}{\mathcal{K}}
\newcommand{\calM}{\mathcal{M}}
\newcommand{\calMCT}{\mathcal{M}^{\ssst{(CT)}}}
\newcommand{\calN}{\mathcal{N}}

\newcommand{\barM}{\bar{\mathcal{M}}}

\newcommand{\topo}{\Omega}
\newcommand{\toposet}[1]{\mathcal{T}_{#1}}
\newcommand{\topodiaset}{\mathcal{G}_{\topo}}
\newcommand{\topoden}{\mathcal{D}_\topo}
\newcommand{\diaset}[1]{{\mathcal{G}}_{\ssst{#1}}}
\newcommand{\singlearg}[1]{
\ifx&#1&
\else
(#1)   
\fi
}

\newcommand{\doublearg}[2]{
\ifx&#2&
(#1)
\else
(#1,#2)   
\fi
}
\newcommand{\ampindices}[5]{{#1}_{{#2,#3}}^{#4 }\singlearg{#5}}
\newcommand{\ratamp}[4]{{\ampindices{\delta \calR}{#1}{#2}{#3}{#4}}}
\newcommand{\deltaZ}[4]{{\ampindices{\delta Z}{#1}{#2}{#3}{#4}}}
\newcommand{\deltaZtilde}[4]{{\ampindices{\delta \tilde Z}{#1}{#2}{#3}{#4}}}

\newcommand{\calV}{\mathcal{V}}
\newcommand{\calW}{\mathcal{W}}
\newcommand{\seg}{S}

\newcommand{\calU}{\mathcal{U}}

\newcommand{\segment}[2]{\seg_{#2}^{(#1)}}

\newcommand{\helisegment}[2]{\heli_{#2}^{(#1)}}

\newcommand{\helic}[1]{\heli^{(#1)}}                   
\newcommand{\helig}{\heli}                             
\newcommand{\helipc}[2]{\helihat_{#2}^{(#1)}}          
\newcommand{\helipcc}[2]{\helicheck_{#2}^{(#1)}}       

\newcommand{\numc}[1]{\calN^{(#1)}}                 

\newcommand{\fullamp}[2]{{\calM}_{{#1},{#2}}}

\newcommand{\fullampbar}[2]{{\barM}_{{#1},{#2}}}
\newcommand{\colfac}[2]{{C}_{{#1},{#2}}}
\newcommand{\vertex}[1]{\calV_{#1}}
\newcommand{\denc}[1]{\mathcal{D}^{(#1)}(\bar q_{#1})} 
\newcommand{\col}{\mathrm{col}}

\newcommand{\re}{\mathrm{Re}}

\newcommand{\redone}{Red1}
\newcommand{\redtwo}{Red2}

\newcommand{\rd}{\mathrm d}

\definecolor{bluemar}{rgb}{0,0,.5}
\definecolor{redmar}{rgb}{.8,0,0}
\definecolor{greenmar}{rgb}{0,.5,0}

\def\Dn{D_{\mathrm{n}}}

\graphicspath{{./figures/}}

\title{Status of two-loop automation in OpenLoops}

\author[a]{N.~Sch\"ar}
\author*[b, a]{M.~F.~Zoller}

\affiliation[a]{PSI Center for Neutron and Muon Sciences, 5232 Villigen PSI, Switzerland}

\affiliation[b]{Universit\"at Z\"urich, CH-8057 Z\"urich, Switzerland}

\emailAdd{max.zoller@physik.uzh.ch}

\abstract{
The calculation of hard scattering amplitudes up to NLO is automated in numerical tools, such as OpenLoops. The LHC and future experiments, however, demand high-precision predictions at NNLO and beyond for a wide range of particle processes. Hence, the development of a fully automated tool for numerical NNLO calculations is an important goal.\\
In order to perform a numerical calculation, we decompose $D$-dimensional two-loop amplitudes into Feynman integrals with four-dimensional numerators and $(D-4)$-dimensional remainders, which contribute to the finite result through the interaction with the poles of Feynman integrals and are reconstructed during the subtraction procedure for these poles from universal rational terms. The integrals with four-dimensional numerators are further decomposed into loop momentum tensor integrals and tensor coefficients.
We present the status of OpenLoops with respect to these building blocks.
The algorithm for the construction of the tensor coefficients is implemented for QED and QCD corrections to the SM in a fully automated way. Recently, the renormalisation procedure and the reconstruction of the interplay of $(D-4)$-dimensional numerator parts with UV poles through two-loop rational counterterms has been implemented and validated using an in-house library for the reduction of simple tensor integrals.}

\FullConference{Loops and Legs in Quantum Field Theory (LL2024)\\
 14-19, April, 2024\\
Wittenberg, Germany\\}

\begin{document}
\maketitle

\section{Introduction}
\label{sec:intro}
Perturbative scattering amplitudes are a key ingredient for Monte Carlo simulations of collider processes. Tree and one-loop amplitudes, required for LO and NLO predictions, have been available from fully automated numerical tools, such as \OpenLoops{} \cite{Cascioli:2011va,Buccioni:2017yxi,Buccioni:2019sur}, for many years.
In order to meet the precision requirements of the LHC and future colliders, NNLO predictions for a wide range of processes are essential, which makes a two-loop version of \OpenLoops{} an important goal.
An automated numerical tool is based on two principles:\\[-5mm]
\begin{itemize}
 \item The {\bf automated} calculation of scattering amplitudes for many processes requires their decomposition into process-independent constituents, from which we can construct them in a recursive way.\\[-8mm]
 \item $D$-dimensional quantities cannot be computed {\bf numerically} in a direct way, but have to be constructed from quantities projected to integer dimensions. We construct integrand numerators in four dimensions and reconstruct $(D-4)$-dimensional parts through process-independent counterterms, allowing also for the automation of this part.\\[-5mm]
\end{itemize}
 Scattering amplitudes at a given loop order $L$ are computed from Feynman diagrams $\Gamma$,
\be
\barM_{L}(\heli) \,=\, \sum\limits_\Gamma \fullampbar{L}{\Gamma}(\heli){}, \label{eq:ampfromdias}
\ee
where $\heli$ denotes the helicity configuration of the external particles of the process at hand and the bar an amplitude in $D$ dimensions.
For processes with $\calM_{0}\neq 0$ the helicity and colour-summed squared tree-level amplitude
\be
\calW_{\ssst{LO}}\,=\,\f{1}{N_{\rm{hcs}}}
\sum\limits_{\heli,\col}
|\calM_{0}(\heli)|^2{},
\label{M2Wtree}
\ee
constitutes the LO contribution of the scattering probability density.
Here, $1/N_{\rm{hcs}}$ encodes the average over initial-state helicity and colour d.o.f as well as symmetry factors 
(see \cite{Buccioni:2019sur}).
A NLO calculation consists of a real-emission contribution, which has the same form as \eqref{M2Wtree} with one extra unresolved particle, and the virtual contribution
computed from the Born-loop interference
\be
\calW_{\ssst{NLO}}^{\ssst{virtual}}\,=\,\f{1}{N_{\rm{hcs}}}
\sum\limits_{\heli,\col}
2\,\re \Big[\calM_{0}^*(\heli)\mathbf{R}\barM_{1}(\heli)\Big],
\label{M2Wone}
\ee
where
\bea
{\textbf{R}}\, \barM_{1} (\heli)
&=&  \calM_1 (\heli) + \calMCT_{0,1}(\heli)
\,.
\label{eq:masterformula1}
\eea
denotes the renormalisation procedure. The first term on the rhs of~\eqref{eq:masterformula1}
is the unrenormalised amplitude computed from Feynman integrals with numerator dimension $\Dn=4$, which can be decomposed into tensor integrals and numerically constructed coefficients. The second term $\calMCT_{0,1}$ stands for the tree-level amplitude with all relevant insertions of one-loop UV counterterms and rational terms \cite{Ossola:2008xq,
Draggiotis:2009yb,
Garzelli:2009is,
Pittau:2011qp}, where the latter reconstruct the $(D-4)$-dimensional numerator parts.

A NNLO calculation consists of a double-real
and a real-virtual part, which have the same form as \eqref{M2Wtree} with two and \eqref{M2Wone} with one extra particle respectively, and the double-virtual contribution
\be
\calW_{\ssst{NNLO}}^{\ssst{virtual}}\,=\,
\sum\limits_{\heli,\col}
2\,\re \Big[\calM_{0}^*(\heli)\,\mathbf{R}\barM_{2}(\heli)\Big] + |\mathbf{R}\barM_{1}(\heli)|^2
\label{M2Wtwobar}
\ee
with
\bea
\mathbf{R}\barM_{2}(\heli) &=&
\calM_{2}(\heli) +
\calMCT_{1,1}(\heli) +
\calMCT_{0,2}(\heli) +
\calMCT_{0,1,1}(\heli){}\,.
\label{eq:masterformula2}
\eea
While the first term on the rhs is the unrenormalised two-loop
amplitude in $\Dn=4$,
each of the three additional contributions
embodies the counterterms for the subtraction of
UV divergences in combination with rational counterterms \cite{Pozzorini:2020hkx,Lang:2020nnl,Lang:2021hnw} for the
reconstruction of the contributions of the $(D-4)$-dimensional numerator
parts. The term $\calMCT_{1,1}(\heli)$ denotes the one-loop amplitude with all relevant
one-loop counterterm insertions, while
$\calMCT_{0,2}(\heli)$ and $\calMCT_{0,1,1}(\heli){}$
correspond, respectively, to the tree-level amplitudes with
single two-loop and double one-loop counterterm insertions. Note that in any renormalisable model a finite set of process-independent UV and rational counterterms exists, allowing for the automation of these terms with an extension of the one-loop \OpenLoops{}.

The amplitudes on the rhs of \eqref{eq:masterformula2} are further decomposed into
loop momentum tensor integrals.

In practice we always compute the helicity and colour-summed interference of the terms in \eqref{eq:masterformula1} and \eqref{eq:masterformula2} with the Born, needed in \eqref{M2Wone} and \eqref{M2Wtwobar} respectively.
The one and two-loop contributions read\footnote{The one-loop contributions with counterterm insertions are decomposed and constructed in the same way as the one-loop contributions. Tree-level contributions with and without counterterm insertions are constructed purely numerically.}
\bea
\sum_{\col,\heli} 2\,\re \Big[\calM_{0}^*(\heli)\,\calM_{1}(\heli)\Big] \!\!\!&=&\!\!\! \sum_{\topo\in\toposet{1}}
\sum_{r_1=0}^{R_1}
{\calU}_{\mu_1 \cdots \mu_{r_1}}^{(\topo)}
\int\!\!\!\rd\momq_1
\f{q_1^{\mu_1}\cdots q_1^{\mu_{r_1}}
   }{\topoden(\momq_1)}, \label{eq:M0M1decomp} \\
\sum_{\col,\heli} 2\,\re \Big[\calM_{0}^*(\heli)\,\calM_{2}(\heli)\Big] \!\!\!&=&\!\!\! \sum_{\topo\in\toposet{2}}
\sum_{r_1=0}^{R_1}\sum_{r_2=0}^{R_2}
{\calU}_{\mu_1 \cdots \mu_{r_1} \nu_1 \cdots \nu_{r_2}}^{(\topo)}
\int\!\!\!\rd\momq_1\!\!\int\!\!\!\rd\momq_2
\f{q_1^{\mu_1}\cdots q_1^{\mu_{r_1}}\,q_2^{\nu_1}\cdots q_2^{\nu_{r_2}}
   }{\topoden(\momq_1,\momq_2)}, \label{eq:M0M2decomp}
\eea
where the integration measure in loop momentum space is defined as
$\int\!\rd\momq_i = \mu^{2\eps}\! \int\!\!\! \f{\rd^{^D}\! \bar
q_i}{(2\pi)^{^D}}$.
The first sum on the rhs is performed over the set of all $L$-loop topologies $\toposet{L}$ of the given process, where a topology is defined by a product $\topoden$ of scalar propagator denominators.
Usually several Feynman diagrams of a process share the same topology $\topo$. Defining $\topodiaset$ to be the set of these diagrams, the coefficients in \eqref{eq:M0M1decomp} and \eqref{eq:M0M2decomp} can be written as the sums of tensor coefficients of individual diagrams,
\be
{\calU}_{\mu_1 \cdots \mu_{r_1}}^{(\topo)} = \sum_{\Gamma\in \topodiaset}
{\calU}_{\mu_1 \cdots \mu_{r_1} }(\Gamma), \qquad\qquad
{\calU}_{\mu_1 \cdots \mu_{r_1} \nu_1 \cdots \nu_{r_2}}^{(\topo)} = \sum_{\Gamma\in \topodiaset}
{\calU}_{\mu_1 \cdots \mu_{r_1} \nu_1 \cdots \nu_{r_2}}(\Gamma). \label{eq:diacoeffdef}
\ee
The advantage of this decomposition is that the tensor coefficients, can be computed numerically, while keeping the analytical loop momentum structure in the tensor integrals. These can then be reduced with analytical or numerical methods or a mixture thereof.

A full two-loop calculation along these lines requires three main building blocks which have different levels of generality. The tensor coefficients are specific to a given process, while the tensor integrals are shared by a whole process class, and the UV and rational counterterms only depend on the given model.
In the following, we will review the status of these building blocks and their implementation in the \OpenLoops{} framework.

\section{Tensor coefficients}\label{sec:tc}

\begin{figure}[t]
\begin{tabular}{ccccc}
  $\vcenter{\hbox{\scalebox{1.}{\includegraphics[width=0.33\textwidth]{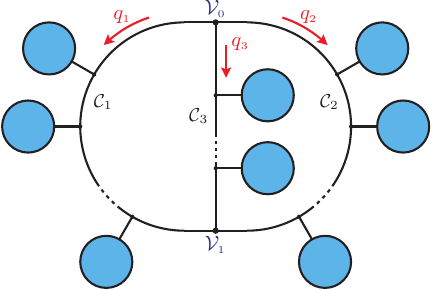}}}}$
& \qquad &
  $\vcenter{\hbox{\scalebox{1.}{\includegraphics[width=0.24\textwidth]{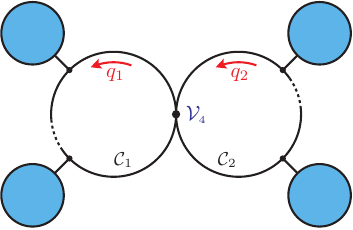}}}}$
& \qquad &
  $\vcenter{\hbox{\scalebox{1.}{\includegraphics[width=0.32\textwidth]{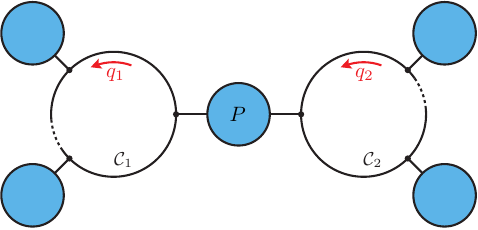}}}}$
\\
(Irred) & & (\redone) & & (\redtwo)
 \end{tabular}
\caption{Categorisation of two-loop diagrams into irreducible (Irred)
and reducible (Red) ones. The latter are further split
into two subcategories, where two one-loop subdiagrams
are either connected to a tree structure $P$ through two vertices (\redtwo)
or are attached to each other through a common quartic vertex (\redone).
The blue blobs denote subtrees connected to internal and external lines. In general $\vertex{0}, \vertex{1}$ can be quartic vertices, in which case an external subtree is also attached there.
\label{fig:twoloopclasses}}
\end{figure}

There are two main categories of two-loop diagrams, irreducible and reducible, as depicted in \reffi{fig:twoloopclasses}.
Irreducible diagrams consist of three propagator chains $\calC_i$, each depending on a single loop momentum $q_i$ $(i=1,2,3)$, with the boundary condition $q_3=-(q_1+q_2)$. These chains are connected by two vertices $\vertex{0},\vertex{1}$, which in general depend on both independent loop momenta $q_1,q_2$.
Reducible diagrams consist of two chains connected by a quartic vertex $\vertex{4}$ or a tree structure $P$, which is either a single propagator or includes external lines. Reducible diagrams can be computed with an extension of the one-loop \OpenLoops{} algorithm, while for diagrams of type Irred a completely new algorithm was developed \cite{Pozzorini:2022ohr}.
In all cases the automated computation of the tensor coefficients exploits the factorisation of diagrams into universal building blocks, which are defined by the Feynman rules and hence only depend on the model.

In \OpenLoops{}, the subtrees $w_a$ needed for the tree and loop diagrams are constructed recursively from smaller subtrees $w_b, w_c$ through steps
\be
w^{\alpha}_a \,=\,
\parbox{0.13\textwidth}{\includegraphics[height=10mm]{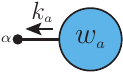}} \,=\,
\parbox{0.13\textwidth}{\includegraphics[height=20mm]{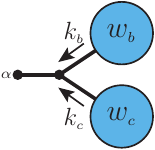}} \,=\,
\f{X_{\beta\gamma}^{\alpha}(k_b,k_c)}{\momk{a}^2-\mass{a}^2}
w^{\beta}_b
w^{\gamma}_c ,
\ee
starting from the wave functions of the external particles. The kernel $X$ is derived from the Feynman rule of the connecting vertex and adjacent propagator with mass $\mass{a}$ and momentum $\momk{a}$.
Tree-level diagrams are constructed by connecting two subtrees into the full diagram.

The amplitude of an L-loop diagram $\Gamma$ is computed as
\be
\calM_{L,\Gamma}(\heli)
=\colfac{L}{\Gamma}\!
\int\!\!\rd\momq_1\cdots\!\int\!\rd\momq_L
\f{\calN(q_1,\cdots,q_L,\heli,\Gamma)
   }{\mathcal{D}_\topo(q_1,\cdots,q_L)},
\ee
with a colour factor $\colfac{L}{\Gamma}$\footnote{Diagrams with non-factorisable colour structures due to quartic vertices are split into colour-factorised contributions, each of which is treated as a separate diagram by \OpenLoops{}.} and a loop integral, of which both the numerator $\calN$ and the denominator $\mathcal{D}_\topo$ have a strongly factorised structure. The denominator is a product of $N$ scalar propagator chains\footnote{For one-loop diagrams $N=1$, for reducible two-loop diagrams $N=2$ and for irreducible two-loop diagrams $N=3$.}
\be
\topoden(\momq_1,\cdots,\momq_L) = \prod\limits_{i=1}^{N}\denc{i} \text{ with }
\ee
each of which is a product of scalar propagator denominators
\bea
\label{eq:dendef}
\denc{i}&=&
D^{(i)}_0(\bar q_i)\cdots
D^{(i)}_{N_i-1}(\bar q_i)\,,
\qquad\mbox{where}\quad
D^{(i)}_a(\bar q_i) \,=\, \left(\bar q_i + p_{ia}\right)^2-m_{ia}^2\,
\eea
with an external momentum $p_{ia}$ and a mass $m_{ia}$.

The numerator of a one-loop diagram
\be
\calN({ q_1},\heli,\Gamma)= {\rm Tr}\left[\seg_1(q_1,\helisegment{1}{1})\!\cdots\!\seg_{N_1}({q_1,\helisegment{1}{N_1}})\right]
\label{eq:oneloopnumdef}
\ee
factorises into loop segments
\bea
\seg_a(q,\helisegment{1}{a})&=&\parbox{0.13\textwidth}{
\includegraphics[height=18mm]{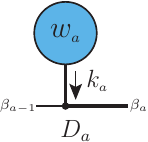}}
=
\{Y_{\sigma}^{a}+ Z_{\nu;\sigma}^{a}\,{q^\nu }
\}\, w^{\sigma}_a(\helisegment{1}{a}) , \label{eq:diasegmentAa}
\eea
each consisting of a loop vertex and propagator encoded in the universal building blocks $Y, Z$ and one or two sub-trees $w_a$ with external momentum $k_a$ and helicity configuration $\helisegment{1}{a}$. Each segment is a matrix with Lorentz or spinor indices $\beta_{a-1}, \beta_a$, and the trace in \eqref{eq:oneloopnumdef} connects the indices $\beta_{0}$ and $\beta_{N}$.
Starting from two loops, diagrams factorise into $N$ chains \be
\numc{i}(q_i,\helic{i}) = \segment{i}{0}({q_i,\helisegment{i}{0}})\cdots\segment{i}{N_i-1}({q_i},\helisegment{i}{N_i-1}).
\ee
corresponding to the denominator chains $\denc{i}$.
Each $\numc{i}$ is a product of loop segments of the same form \eqref{eq:diasegmentAa} as at one-loop level.
The helicity labels are defined in an additive way (see \cite{Pozzorini:2022ohr}), such that the helicity configuration of a chain is given by $\helic{i}=\helisegment{i}{1}+\ldots+\helisegment{i}{N_i-1}$.
The open Lorentz or spinor indices of the numerator chains are connected to vertices $\vertex{i}$ or tree structures $P$. At two loops the numerator factorisation reads\footnote{The helicity label $\helisegment{V}{j}$ appears if $\vertex{j}$ is a quartic vertex with an attached subtree. $\helisegment{P}{}$ is the helicity of the connecting tree structure of a \redtwo{} diagram. The global helicity label $\helig$ is the sum of these labels and the chain helicities $\helic{i}$.}
\bea
\calN({ q_1, q_2},\heli,\Gamma)= \begin{cases}
                                  \prod\limits_{i=1}^{3}\numc{i}({ q_i},\helic{i})
                                  \prod\limits_{j=0}^{1}\vertex{j}({ q_1},{ q_2},\helisegment{V}{j})
                                  & \text{ for } \Gamma \in \diaset{Irred},\\
                                  \prod\limits_{i=1}^{2}\numc{i}({ q_i},\helic{i})
                                  \vertex{4}
                                  & \text{ for } \Gamma \in \diaset{Red1},\\
                                  \prod\limits_{i=1}^{2}{\rm Tr}\left[\numc{i}({ q_i},\helic{i})\right]
                                  P(\helisegment{P}{})
                                  & \text{ for } \Gamma \in \diaset{Red2},
                                 \end{cases}
\label{eq:A2d}
\eea
defining the sets of all two-loop diagrams $\diaset{c}$ of the three categories depicted in Fig.~\ref{fig:twoloopclasses}.

Multiplying the numerator of a diagram $\Gamma$ with the corresponding Born-colour interference
\be
\calU_0(h,\Gamma)
 = 2\,\sum_{\col}\calM^*_0(\helig)\,\colfac{L}{\Gamma},
\label{eq:colborn_two}
\ee
yields its contribution to the squared matrix element \eqref{M2Wone} or \eqref{M2Wtwobar}. The tensor coefficients
of this object as defined in \eqref{eq:diacoeffdef} are computed from universal building blocks in recursion steps
of the form
\be
\hat\calU_{n}=\hat\calU_{n-1} \cdot \calK_{n}. \label{eq:tcrec_general}
\ee
The building blocks $\calK_{n}$ are the loop segments \eqref{eq:diasegmentAa} of the loop chains, the connecting structures $\vertex{j}, P$ in \eqref{eq:A2d} and the factor $\calU_0$ defined in \eqref{eq:colborn_two}, all of which
consist of a process-independent kernel and/or a previously computed tree structures.
At one-loop level the recursion has the form \cite{Buccioni:2017yxi,Buccioni:2019sur}
\be
\calU_{n}(q_1,\helipcc{1}{n})=\sum_{\helisegment{1}{n}} \calU_{n-1}(q_1,\helipcc{1}{n-1}) \cdot \seg_{n}(q_1,\helisegment{1}{n})
\quad
\text{with}\quad
{ \helipcc{1}{n}= \helig -\sum\limits_{a=1}^n\helisegment{1}{a}}, \label{eq:tcrec_one_helsum}
\ee
with steps $n=1,\ldots,N_1$ and initial condition \eqref{eq:colborn_two}. Exploiting the fact that only $\calU_{0}$ and $\seg_{n}$ depend on the helicity dof of the external particles in the n-th segment, we sum over these helicities on the fly (see \cite{Buccioni:2017yxi} for details).
Since the loop momenta on the two chains of reducible two-loop diagrams are independent and the connecting structures $\vertex{4}$ or $P$ do not depend on any loop momentum, the calculation can be factorised into two one-loop recursions,
one of the form \eqref{eq:tcrec_one_helsum} and one of the form
\be
\calN_{n}(q_2,{ \helipc{2}{n}}) = \calN_{n-1}({ q_2},{ \helipc{2}{n-1}})\cdot\segment{2}{n}({ q_2},{ \helisegment{2}{n}})
 \quad\text{with} \quad
 \helipc{2}{n} = \sum\limits_{a=1}^{n} \helisegment{2}{a}
  \quad\text{and} \quad \calN_{-1}=\idop
 \label{eq:tcrec_one_wohelsum}
\ee
Contracting the tensor coefficients computed in the recursion \eqref{eq:tcrec_one_wohelsum} for chain $\calC_2$ with the corresponding pre-computed one-loop tensor integrals and either $\vertex{4}$ or $P$, results in an object that can be treated like an external subtree in a segment of the the recursion \eqref{eq:tcrec_one_helsum} used on $\calC_1$.

For irreducible diagrams we developed a completely general and highly efficient recursive algorithm, described in detail in \cite{Pozzorini:2022ohr} and summarized in \cite{Zoller:2022ewt}. Here the chains are sorted by number of segments such that $N_1\geq N_2\geq N_3$ and the vertices $\vertex{0,1}$ by vertex type, which plays an important role for the CPU efficiency of the algorithm. Then the numerator $\numc{3}({ q_3},\helic{3})$ of the shortest chain is constructed first with an algorithm of the form \eqref{eq:tcrec_one_wohelsum} and used as a single building block in the following recursion of the form \eqref{eq:tcrec_general} with the ordered set of building blocks
\be
\calK_{n} \in \left\{\calU_0,\seg^{(1)}_1,\ldots,\seg^{(1)}_{N_1-1},\vertex{1},\numc{3},\vertex{0},\seg^{(2)}_1,\ldots,\seg^{(2)}_{N_1-1} \right\}.
\ee
Here, we employ an on-the-fly helicity summation as introduced in \eqref{eq:tcrec_one_helsum}, reducing the number of helicity dof of the computed coefficients in every step.
This procedure balances the high tensor rank complexity in later steps of the algorithm with a low number of helicities, for which the intermediate tensor coefficients are computed.
The algorithms for all categories of two-loop diagrams are fully implemented and validated for QED and QCD corrections to the Standard Model.

\section{Tensor integrals - First steps} \label{sec:ti}
At one-loop level tensor integrals are reduced on the fly \cite{Buccioni:2017yxi}, i.e.~during the tensor coefficient construction, to a small set of scalar integrals, which are then evaluated with \Collier{}  \cite{Denner:2016kdg} or \OneLoop{} \cite{vanHameren:2010cp}. Alternatively, \Collier{} can be used for the full reduction and evaluation of tensor integrals. This is required for the pre-computed tensor integrals used in the construction of reducible two-loop diagrams or for one-loop squared amplitudes.

In order to validate the full two-loop \OpenLoops{} framework, in particular the implementation of UV and rational counterterms described in the next section, we implemented a first tool for the reduction of simple tensor integrals to scalar master integrals based on the method of projectors and IBP reduction \cite{ChetyrkinMINCER}. This method is general, but for higher tensor ranks, high-point topologies, and different propagator masses, the intermediate results become very large limiting the practical scope of this approach.

The tool consists of general \Fortran{} part with all numerical routines and
an analytical generator written in \Mathematica{}, which for any tensor integral
\be
I^{\mu_1\cdots\mu_r\nu_1\cdots\nu_s} =
\int\!\!\!\rd\momq_1\!\!\int\!\!\!\rd\momq_2
\f{q_1^{\mu_1}\cdots q_1^{\mu_{r_1}}\,q_2^{\nu_1}\cdots q_2^{\nu_{r_2}}}{\topoden(\momq_1,\momq_2)}
\ee
of a given two-loop topology $\Omega$ performs the following steps:
 First, we derive a covariant decomposition of the final result,
 \be
 I^{\mu_1\cdots\mu_r\nu_1\cdots\nu_s} = \sum_{\tau=1}^{N_t} T_\tau^{\mu_1\cdots\mu_r\nu_1\cdots\nu_s} I_\tau , \label{eq:covdecomptwo}
 \ee
 where we sum over all $N_t$ possible tensor structures $T_\tau$ constructed from the metric tensor and the external momenta. This yields a system of equations ($\alpha=1,\ldots,N_t$)
 \be
 T_{\alpha, \mu_1\cdots\mu_r\nu_1\cdots\nu_s} I^{\mu_1\cdots\mu_r\nu_1\cdots\nu_s} = \sum_{\tau} M_{\alpha\tau} I_\tau
 \quad \text{with}\; M_{\alpha\tau}=T_{\alpha, \mu_1\cdots\mu_r\nu_1\cdots\nu_s}T_\tau^{\mu_1\cdots\mu_r\nu_1\cdots\nu_s}, \label{eq:projcovdecomptwo}
 \ee
 which is solved for the coefficients $I_\tau$ by inverting the matrix $M_{\alpha\tau}$. Since the lhs of \eqref{eq:projcovdecomptwo} is a scalar integral, the $I_\tau$ are now expressed as linear combinations of scalar integrals. These are reduced to master integrals with \FIRE{} \cite{Smirnov:2019qkx,Smirnov:2023yhb} followed by an expansion in $\eps=(4-D)/2$ to the required order. The final expressions for the tensor integrals in terms of the master integrals are then stored into a \Fortran{} library for a given topology, which is further structured into modules for specific tensor ranks. These libraries can then be called by the general \Fortran{} part.
For the evaluation of the master integrals we use a combination of analytical expressions -- currently massless two and three-point integrals \cite{Birthwright:2004kk} -- implemented in our code and the numerical tool \FIESTA{} \cite{Smirnov:2021rhf}. The generator part is written in a fully general way, but for memory and CPU performance reasons we currently only used this tool for the full set of massless two-point and three-point tensor integrals appearing in renormalisable theories, as well as a few four-point ones. This is sufficient for the validation of the UV and rational counterterms and the re-computation of some simple amplitudes, such as the QED and QCD vertex corrections. It is also expected to be useful for the  validation and completion of a more powerful new method and tool for higher-point topologies and higher tensor ranks which is currently being developed.

\section{Renormalisation and UV rational terms - Implementation and validation} \label{sec:rt}

\newcommand{\diaheightCT}{2.9cm}
In \cite{Pozzorini:2020hkx} we derived a formula for the computation of renormalised two-loop amplitudes in $D$ dimensions, which is compatible with the above construction of tensor coefficients and tensor integrals and for a two-loop diagram $\Gamma$ of type Irred or \redone{} reads\footnote{These are the diagram types which become 1PI upon amputation of the external subtrees.}
\be
{\textbf{R}}\, \fullampbar{2}{\Gamma}
\,=\, \fullamp{2}{\Gamma}\,+\,
\sum  \limits_{\gamma} \lb \deltaZ{1}{\gamma}{}{} +\deltaZtilde{1}{\gamma}{}{} + \ratamp{1}{\gamma}{}{} \rb \cdot \fullamp{1}{\Gamma/\gamma}
\,+\, \lb
\deltaZ{2}{\Gamma}{}{} + \ratamp{2}{\Gamma}{}{} \rb \fullamp{0}{\Gamma}
\,.
\label{eq:masterformula2dia}
\ee
While the lhs is fully $D$-dimensional, the amplitudes on the rhs are computed with four-dimensional loop numerators and $D$-dimensional denominators. The sum in the second term is performed over all UV-divergent one-loop subdiagrams $\gamma\subset\Gamma$ and $\fullamp{1}{\Gamma/\gamma}$ the one-loop amplitude resulting from contracting $\gamma$ to a vertex in $\Gamma$. The UV counterterms $\deltaZ{1}{\gamma}{}{}$ and $\deltaZtilde{1}{\gamma}{}{}$ subtract the subdivergences\footnote{$\deltaZ{1}{\gamma}{}{}$ is the usual UV counterterm with its tensor structure projected from $D$ to four dimensions, while $\deltaZ{1}{\gamma}{}{}$ is a new but also universal counterterm which is non-zero only for one-loop subdiagrams of mass dimension $2$.} stemming from $\gamma$, while the one-loop rational term $\ratamp{1}{\gamma}{}{}$  \cite{Ossola:2008xq,
Draggiotis:2009yb,
Garzelli:2009is,
Pittau:2011qp} restores the interplay of $(D-4)$-dimensional numerator parts of the subdiagram amplitude $\fullamp{1}{\gamma}$ with its UV divergence. The last term consists of the tree-level amplitude derived from contracting the usual two-loop UV counterterm $\deltaZ{2}{\Gamma}{}{}$ projected to four dimensions and a two-loop rational term restoring the remaining $(D-4)$-dimensional numerator parts of $\fullampbar{2}{\Gamma}$ interacting with UV poles of the loop integral. It is straightforward to extend $\gamma$ and $\Gamma$ in \eqref{eq:masterformula2dia} from single diagrams to vertex and propagator functions due to the linearity of $\textbf{R}$, and \eqref{eq:masterformula2dia} to \eqref{eq:masterformula2} due to the linearity of \eqref{eq:ampfromdias}. The full set of two-loop rational terms for QED and QCD corrections to the Standard Model has been presented in \cite{Lang:2020nnl,Lang:2021hnw}. While this procedure recovers the full contribution of the UV poles the interplay of $(D-4)$-dimensional numerator parts with IR poles is still under investigation, but first insights were presented in \cite{Zhang:2022rft}.

\begin{figure}[t]
\begin{tabular}{ccccc}
  $\vcenter{\hbox{\scalebox{1.}{\includegraphics[height=\diaheightCT]{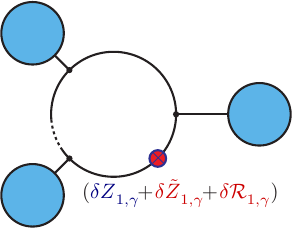}}}}$
& \quad\qquad &
  $\vcenter{\hbox{\scalebox{1.}{\includegraphics[height=\diaheightCT]{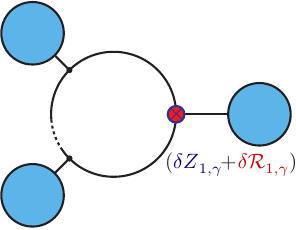}}}}$
& \quad\qquad &
  $\vcenter{\hbox{\scalebox{1.}{\includegraphics[height=\diaheightCT]{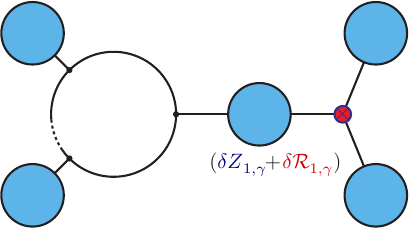}}}}$
\\
(O1a) & & (O1b) & & (O2)
 \end{tabular}
\caption{Categories of one-loop diagrams with counterterm insertions. The master formula for two-loop diagrams of type Irred requires up to three contributions of type O1a and O1b. Two-loop \redone{} diagrams each require up to two O1b contributions, \redtwo{} diagrams each require up to two contributions of type O2.\\[-5mm]
\label{fig:oneloopCTclasses}}
\end{figure}

\begin{figure}[t]
\begin{tabular}{cccc} \qquad\qquad\qquad\qquad &
  $\vcenter{\hbox{\scalebox{1.}{\includegraphics[height=\diaheightCT]{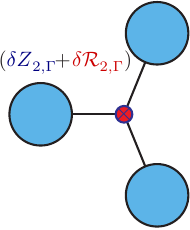}}}}$
& \qquad\qquad\qquad &
  $\vcenter{\hbox{\scalebox{1.}{\includegraphics[height=\diaheightCT]{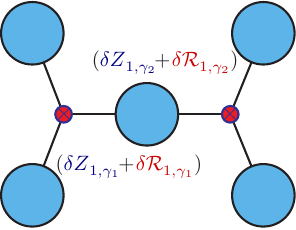}}}}$
\\
& (T1) & & (T2)
 \end{tabular}
\caption{Categories of tree-level diagrams with counterterm insertions. The master formula for two-loop diagrams of type Irred and \redone{} with a global UV divergence require a contribution of type T1, while two-loop diagrams of type \redtwo{} require a contribution of type T2 if both subdiagrams are UV divergent.
\label{fig:treeCTclasses}}
\end{figure}

For a reducible diagram $\Gamma$ of type \redtwo{}, which factorises into one-loop subdiagrams $\gamma_1,\gamma_2$ to which the one-loop procedure can be applied, the master formula reads
\be
{\textbf{R}}\, \fullampbar{2}{\Gamma}
\,=\,  \fullamp{2}{\Gamma}\,+\,
\sum  \limits_{i=1}^2 \lb \deltaZ{1}{\gamma_i}{}{} + \ratamp{1}{\gamma_i}{}{} \rb \cdot \fullamp{1}{\Gamma/\gamma_i}
\,+\,
\prod  \limits_{i=1}^2 \lb \deltaZ{1}{\gamma_i}{}{} + \ratamp{1}{\gamma_i}{}{} \rb \fullamp{0}{\Gamma/(\gamma_1\cup\gamma_2)}\,.
\label{eq:masterformula2diared2}
\ee
The relevant types of one-loop and tree-level diagrams with counterterm insertions for a full two-loop calculation are depicted in  \reffi{fig:oneloopCTclasses} and \reffi{fig:treeCTclasses}, respectively.

The computation of all relevant contributions at two-loop, one-loop and tree level with the proper counterterm insertions
as well as their combination to fully renormalised $D$-dimensional amplitudes are implemented in the \OpenLoops{} framework. The counterterms for QED and QCD are currently computed in the $\msbar$ scheme, but this is simple to extend to other schemes \cite{Lang:2020nnl}.
There are some technical subtleties in extending \OpenLoops{} for diagrams of type O1a and O1b, such as the interplay of the $\f{1}{\eps}$ pole of $\deltaZ{1}{\gamma}{}{}$ with the $\mathcal{O}(\eps)$ contributions of the one-loop integral.
Since these are not provided by our external one-loop integral tools \cite{Denner:2016kdg, vanHameren:2010cp}, we currently
perform the one-loop tensor integral reduction and evaluation with the tool described in \refse{sec:ti}. The same applies to integrals with squared scalar propagators stemming from diagrams of type O1b.

In order to validate this implementation we compute full two-loop off-shell vertex functions, which avoid IR divergences, in QED and QCD with \OpenLoops{}, using the building blocks described in the previous sections. As a first step, we checked the cancellation of the UV poles in \eqref{eq:masterformula2} for several two and three-point amplitudes in QED and QCD, which is non-trivial since, in addition to the loop amplitudes and UV counterterms, also $\ratamp{2}{\Gamma}{}{}$ usually exhibits a $\f{1}{\eps}$ pole. The check for the four-point QCD vertex function and the calculation of the finite parts of all QED and QCD vertex functions, to be compared to the literature, are ongoing and will serve as a full validation of our renormalisation procedure as well as a first application of two-loop rational terms in \OpenLoops{}.

\section{Conclusion}
We presented the status of an automated numerical two-loop tool in the \OpenLoops{} framework consisting of three main building blocks. For the construction of the two-loop tensor coefficients a completely general and highly efficient algorithm has been developed and implemented. The reduction of the corresponding tensor integrals has been implemented for simple cases with the scope of testing the full two-loop framework. The complete renormalisation procedure including the two-loop rational terms has been implemented. A more powerful method for the tensor integral reduction and a consistent framework for the treatment of two-loop rational terms of IR origin are under development, which will open the door to a wide range of applications for this tool.

\section*{Acknowledgements}
This research was supported by the Swiss National Science Foundation (SNSF) through contracts
PZ00P2-179877 and TMSGI2-211209.
\providecommand{\href}[2]{#2}\begingroup\raggedright\endgroup

\end{document}